\documentclass[
  reprint,
  superscriptaddress,
  amsmath,amssymb,
  aps,
  prl
]{revtex4-2}

\usepackage{graphicx}
\usepackage{dcolumn}
\usepackage{bm}

\usepackage{mathrsfs}
\usepackage[ruled,noline,vlined]{algorithm2e}
\SetKwInOut{KwInit}{Initialize}

\newlength{\InLabel}
\newlength{\OutLabel}
\AtBeginDocument{
  \settowidth{\InLabel}{\textbf{Input: }}
  \settowidth{\OutLabel}{\textbf{Output: }}
}
\newcommand{\KwInH}[1]{\KwIn{\hangafter=1\hangindent=\InLabel #1}}
\newcommand{\KwOutH}[1]{\KwOut{\hangafter=1\hangindent=\OutLabel #1}}
\newcommand{\HKwBreak}{\\\hspace*{\InLabel}}

\begin{document}

\preprint{APS/123-QED}

\title{Dual-wavelength Fourier Ptychographic Topography}

\author{Yi~Shen}
\email{yishen@bu.edu}
\affiliation{Computational Imaging Systems Lab, Department of Electrical and Computer Engineering,
Boston University, Boston, MA 02215, USA}

\author{Tongyu~Li}
\affiliation{Computational Imaging Systems Lab, Department of Electrical and Computer Engineering,
Boston University, Boston, MA 02215, USA}

\author{Hao~Wang}
\affiliation{Computational Imaging Systems Lab, Department of Electrical and Computer Engineering,
Boston University, Boston, MA 02215, USA}

\author{Jinyong~Kim}
\affiliation{Advanced Process Development Lab, Korea}

\author{Hojun~Lee}
\affiliation{Advanced Process Development Lab, Korea}

\author{Wookrae~Kim}
\affiliation{Advanced Process Development Lab, Korea}

\author{Jonghyeok~Park}
\affiliation{Advanced Process Development Lab, Korea}

\author{Junho~Shin}
\affiliation{Core Technology R\&D Team, Samsung Electronics Co., Ltd., Korea}

\author{Seungbeam~Park}
\affiliation{Core Technology R\&D Team, Samsung Electronics Co., Ltd., Korea}

\author{Lei~Tian}
\email{leitian@bu.edu}
\affiliation{Computational Imaging Systems Lab, Department of Electrical and Computer Engineering,
Boston University, Boston, MA 02215, USA}

\date{\today}

\begin{abstract}
We introduce a dual-wavelength Fourier ptychographic topography (FPT) method that extends the $\lambda/2$ height-range limit of single-wavelength FPT. By reconstructing complex fields at two illumination wavelengths and exploiting their phase difference, the method achieves an effective synthetic wavelength $\lambda_s$ and an unambiguous range of $\lambda_s/2$ without reducing lateral resolution. A noise-robust wrapped-number search is used to select per-pixel integer pairs $(k_1,k_2)$, and a global refinement with circular TV regularization and soft bounds improves stability and preserves height discontinuities.
The approach is validated through rigorous scattering-model–based simulations and experiments on structured silicon samples, demonstrating accurate height recovery in regimes where single-wavelength FPT exhibits phase wrapping. We analyze the limits of the FPT forward model and identify aspect ratio (AR) and a phase modulation transfer function (ph-MTF) as key predictors of reconstruction fidelity. Simulations and experiments show that increasing AR beyond a practical threshold causes loss of high-frequency phase transfer and destabilizes dual-wavelength unwrapping. Within this AR range, dual-wavelength FPT provides robust, high-resolution topography suitable for semiconductor and industrial metrology.
\begin{description}
\item[Keywords]
Computational imaging, Fourier ptychography, Optical surface metrology, \\
Phase unwrapping, Semiconductor inspection
\end{description}
\end{abstract}

\maketitle

\section{Introduction}
Surface topography techniques providing wide field-of-view (FOV), high spatial resolution, and nanometer-level accuracy, play  roles across numerous scientific and industrial fields, including semiconductor metrology, material characterization, and advanced manufacturing processes~\cite{poon1995comparison, assender2002topography, degroot2015interference}. 
Conventional topography measurements rely on scanning probe methods, such as atomic force microscopy (AFM)~\cite{orji2018metrology}, or optical interferometric techniques, including white-light interferometry and phase-shifting interferometry~\cite{creath1988poi, degroot2015interference}. 
Although these techniques offer high accuracy, they are often hindered by intrinsic limitations: AFM suffers from slow scanning speeds, mechanical fragility, and potential sample damage, whereas traditional optical interferometry approaches, while non-contact and faster, are susceptible to environmental disturbances and have limited measurement ranges.

Recently, Fourier ptychography (FP) has emerged as a promising computational imaging modality that synthesizes multiple low-resolution intensity images captured under varying illumination conditions to reconstruct high-resolution amplitude and phase information~\cite{zheng2013fpm,tian2014multiplexed, choi2018dwfpm, zheng2021review}. 
As a synthetic aperture technique, the ultimate resolution attainable by FP is determined by the effective synthetic numerical aperture (NA) of the system, which corresponds to the sum of the illumination and imaging NAs.
Originally developed in transmission geometry~\cite{zheng2013fpm}, FP was later extended to reflective mode for surface topography, termed Fourier ptychographic topography (FPT)~\cite{guo2015fpmMulti, lee2019reflectiveFPM, wang2023fpt}. 
FPT combines the advantages of wide FOV, high spatial resolution, and computational robustness, effectively addressing many limitations of traditional metrology techniques. 
Nevertheless, single-wavelength FPT is fundamentally limited by phase wrapping, which restricts the unambiguous height range to half the illumination wavelength ($\lambda/2$)~\cite{choi2018dwfpm, zhou2022reviewdw}. 
This constraint severely hampers the applicability of single-wavelength FPT in practical scenarios involving larger topographic variations.

To address the limited height range of single-wavelength illumination, recent studies have explored dual- and multi-wavelength strategies, which substantially extend the measurable range~\cite{creath1988poi, schmit1995extended,  gao2022prif}. In the dual-wavelength case, phase differences from two distinct illumination wavelengths generate an effective synthetic wavelength, $\lambda_s = \lambda_1 \lambda_2 / |\lambda_1 - \lambda_2|$, extending the unambiguous phase range to $\lambda_s/2$. Dual-wavelength FPT leverages this principle by combining phase reconstructions at two wavelengths and resolving height ambiguities through dual-wavelength unwrapping algorithms. However, the direct phase-differencing approach amplifies noise by a factor of $\lambda_s/|\lambda_1 - \lambda_2|$, which can severely degrade accuracy when the wavelengths are closely spaced~\cite{zhou2022reviewdw}. 

A central challenge in dual-wavelength FPT is therefore robust phase unwrapping in the presence of noise and limited single-wavelength reconstruction quality. Existing strategies, such as branch-cut, quality-guided/least-squares, and global graph-cut/PUMA formulations, seek spatially consistent unwrapped solutions, often incorporating total variation (TV) regularization to suppress noise and artifacts while preserving phase steps and edges~\cite{goldstein1988two,bioucas2007puma,rudin1992rof,chambolle2004tv, Kamilov:15}. 
In addition, reflective systems require accurate modeling of defocus across wavelengths. Longitudinal chromatic aberration induces wavelength-dependent focal-plane mismatch that must be corrected to avoid artifacts in cross-wavelength fusion. Digital refocusing in FP provides an effective means to compensate for this axial mismatch without mechanical focal scanning~\cite{horstmeyer2015pathology, zheng2011ledrefocus, zheng2013fpm}. 

Motivated by these challenges, we develop a dual-wavelength FPT approach that overcomes phase-wrapping and chromatic mismatch while mitigating noise amplification, thereby enabling robust topography beyond the $\lambda/2$ limit without sacrificing the lateral resolution set by the synthetic NA.
Our method integrates a per-pixel wrapped-number search, which enforces cross-wavelength height consistency and thereby mitigates noise amplification inherent to naive phase differencing, with a digital refocusing procedure that co-registers axial planes across wavelengths.
To further enhance spatial fidelity, we incorporate a global refinement step that combines phase-wrapping–aware circular TV regularization with soft bound constraints, suppressing noise while preserving true height discontinuities.

We validate the proposed dual-wavelength FPT framework using both rigorous numerical simulations based on a Modified Born Series (MBS) forward model~\cite{li2025rDTreflect} and experimental measurements on structured silicon samples. Across these evaluations, the method reliably recovers accurate and spatially coherent surface profiles in regimes where single-wavelength FPT fails due to phase wrapping. We further quantify the performance limits of FPT by introducing metrics such as the structural aspect ratio (AR) and phase modulation transfer function (ph-MTF). Through the detailed error studies, we show how height, feature width and illumination NA jointly govern the recoverable lateral frequencies and the height accuracy of the proposed technique.

In summary, this work makes four key contributions:
(i) a defocus-aware FPT forward model to achieve cross-wavelength axial co-registration without mechanical adjustment;
(ii) a unified, noise-robust dual-wavelength unwrapping algorithm that enforces cross-wavelength consistency and incorporates a global refinement to preserve true height transitions while improving spatial fidelity;
(iii) a comprehensive performance characterization that quantifies how structural geometry and system parameters jointly determine the practical operating limits and recoverable frequency content of FPT in metrology and inspection applications.

\begin{figure*}[htbp]
  \centering
  \includegraphics[width=0.95\textwidth]{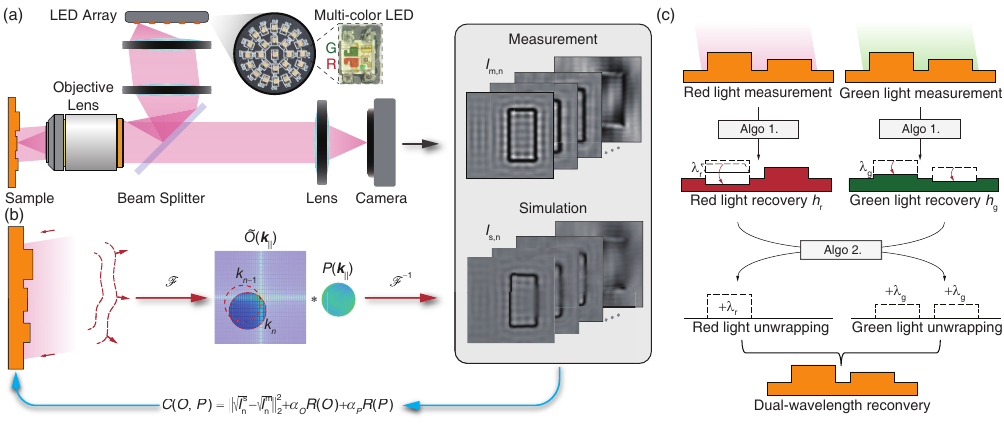}
\caption{
Dual-wavelength FPT pipeline.
(a) The imaging system captures intensity measurements $I_{m,n}$ under programmable angular illumination using red and green LEDs. The object $O(x,y)$ and pupil function $P(u,v)$ are jointly reconstructed by minimizing the mismatch between the measured images $I_{m,n}$ and the simulated images $I_{s,n}$. Gradients $\partial L/\partial O$ and $\partial L/\partial P$ are used for alternating updates. Each wavelength $i$ produces a wrapped phase map $\hat{\varphi}_{i}$. 
(b) Dual-wavelength phase unwrapping algorithm. Sequential red and green illuminations generate wrapped phase maps $\hat{\varphi}_{r}$ and $\hat{\varphi}_{g}$ with Algorithm~\ref{alg:FPT}, each corresponding to multiple candidate height values $h_{k_r}$ and $h_{k_g}$ due to the $2\pi$ ambiguity. By evaluating all candidate index pairs $(k_r, k_g)$ at each pixel, a consistent pair $(k_r', k_g')$ is selected to enable robust unwrapping and height recovery across wavelengths, as shown in Algorithm~\ref{alg:dwFPT}.}
  \label{fig:Pipeline}
\end{figure*}

\section{Theory and Method}
\subsection{Overview of Dual-Wavelength FPT}

In FPT, the object is illuminated sequentially by different LEDs at varying angles in an epi-geometry [Fig.~\ref{fig:Pipeline}(a)]. 
The resulting intensity measurements are used in a ptychographic phase retrieval algorithm to reconstruct the object’s phase map at each  wavelength $\lambda_i$. 
For reflective samples, the recovered phase $\varphi_{i}(\mathbf{r})$ is related to the surface height $h(\mathbf{r})$ by~\cite{wang2023fpt}
\begin{equation}
\varphi_{i}(\mathbf{r}) = \frac{4\pi}{\lambda_i} h(\mathbf{r}).
\label{eq:phase2height}
\end{equation}

A fundamental limitation arises because the recovered phase $\varphi_{i}$ is inherently wrapped into $(-\pi, \pi]$, restricting the unambiguous height range to $\lambda_i/2$. As a result, surface variations exceeding this range are aliased into a reduced interval. While single-wavelength phase unwrapping algorithms exist, their accuracy is fundamentally constrained under noise and reconstruction artifacts~\cite{zhou2022reviewdw}.   

To extend the measurement range, optical metrology often employs two illumination wavelengths to recover independent wrapped phases $\varphi_1$ and $\varphi_2$. Their difference yields a synthetic phase $\varphi_s(\mathbf{r})$ corresponding to a synthetic wavelength $\lambda_s$:  
\begin{equation}
\varphi_s(\mathbf{r}) = \varphi_1(\mathbf{r}) - \varphi_2(\mathbf{r})
= \frac{4\pi}{\lambda_s} h(\mathbf{r}),
\label{eq:synphase}
\end{equation}
where
\begin{equation}
\lambda_s = \frac{\lambda_1 \lambda_2}{|\lambda_1 - \lambda_2|}.
\label{eq:synlambda}
\end{equation}
The synthetic wavelength $\lambda_s$ extends the unambiguous height range by a factor of $\lambda_{2,1} / |\lambda_1 - \lambda_2|$. Accordingly, Eq.~\eqref{eq:synphase} enables recovery of taller structures by resolving phase ambiguities within the half-synthetic-wavelength range, thereby substantially increasing the measurable height range.  

However, the direct phase-differencing method in Eq.~\eqref{eq:synphase} also amplifies noise. Specifically, phase noise in $\varphi_1$ or $\varphi_2$ is magnified in $\varphi_s$ by a factor of $\lambda_{2,1}/|\lambda_1 - \lambda_2|$. If the phase at each wavelength contains an error $\Delta\varphi$, the resulting height error is up to
\begin{equation}
\Delta h = \frac{2\Delta \varphi}{4\pi} \lambda_s ,
\label{eq:synphase_noise}
\end{equation}
corresponding to an amplification relative to the single-wavelength case by a factor of $2\lambda_s/\lambda_{1,2}$. This amplification can significantly degrade reconstruction accuracy in dual-wavelength configurations.  

To address this limitation, we propose a dual-wavelength FPT framework that preserves the extended measurement range while mitigating noise amplification. The framework comprises two main components. First, phase maps at both wavelengths are reconstructed using an improved single-wavelength FPT algorithm [Fig.~\ref{fig:Pipeline}(b)], which integrates (i) wrapped-phase TV regularization to suppress phase retrieval artifacts and (ii) digital refocusing to achieve axial alignment across wavelengths. Second, the framework introduces an optimization-based, noise-robust dual-wavelength phase unwrapping algorithm [Fig.~\ref{fig:Pipeline}(c)].

\subsection{FPT Forward Model}

The FPT measurement under illumination from the $n$-th LED at wavelength $\lambda_i$ 
is modeled as
\begin{equation}
I^s_{n,\lambda_i}(\mathbf{r})
= \Big\lvert \mathscr{F}^{-1}\!\left\{
    \mathscr{F}\!\left[ O_{\lambda_i}(\mathbf{r}) \, S_{\lambda_i,n}(\mathbf{r}) \right]
    P(\mathbf{k})
  \right\} \Big\rvert^{2},
\label{eq:forward}
\end{equation}
where the object field is 
$O_{\lambda_i}(\mathbf{r}) = a(\mathbf{r}) \, e^{\,i\varphi_{i}(\mathbf{r})}$, 
with $a(\mathbf{r})$ denoting the intensity reflected by the sample and 
$\varphi_{i}(\mathbf{r})$ 
the height-induced propagation phase. 
Here, $S_{\lambda_i,n}(\mathbf{r}) = e^{i\,\mathbf{k}_n\cdot\mathbf{r}}$ is the plane-wave illumination from the $n$-th LED, 
$\mathbf{k}_n = \big(2\pi\sin\theta_{x_n}/\lambda_i, 2\pi\sin\theta_{y_n}/\lambda_i\big)$ 
is the corresponding spatial frequency defined by LED angle $(\theta_{x_n},\theta_{y_n})$, 
and $P(\mathbf{k})$ is the pupil function in the Fourier domain $\mathbf{k}=(k_x,k_y)$. 
$\mathscr{F}$ and $\mathscr{F}^{-1}$ denote the 2D Fourier and inverse Fourier transforms, 
with $\mathscr{F}[f(\mathbf{r})](\mathbf{k}) = \int\!\!\!\int f(\mathbf{r})\, e^{-i\mathbf{k}\cdot\mathbf{r}}\, \mathrm{d}\mathbf{r}$~\cite{wang2023fpt}.
The superscript $s$ denotes the simulated intensity from the forward model.

\subsection{Single-Wavelength FPT Algorithm}

\subsubsection{Overall algorithm structure}

The single-wavelength FPT reconstruction is formulated as an optimization problem in which 
the object field $O(\mathbf{r})$ and pupil function $P(\mathbf{k})$ are recovered by minimizing
\begin{equation}
C(O,P) = L(O,P;I^m) + R(O) + R(P),
\end{equation}
where $I^m = \{ I^m_n(\mathbf{r}) \}_{n=1}^{N}$ denotes the measured intensities, 
$L$ is the data-fidelity term, and $R(\cdot)$ denotes the corresponding regularization term for the object and pupil function.  

The data-fidelity term enforces consistency between measured and simulated intensities:
\begin{equation}
L(O,P;I^m) =
\sum_{n=1}^{N} 
\left\| \sqrt{I^s_n(\mathbf{r})} - \sqrt{I^m_n(\mathbf{r})} \right\|_2^2,
\label{eq:cost_intensity}
\end{equation}
where $N$ denotes the total number of measurements.

The regularization term stabilizes the inverse problem by combining $L_2$ and TV penalties:  
\begin{equation}
R(A) = \alpha_\mathrm{A}\|A\|_2^2 
+ \beta_{\mathrm{A,ampl}}\|\,|A|\,\|_{\mathrm{TV}} 
+ \beta_{\mathrm{A,arg}}\|\arg(A)\|_{\mathrm{cTV}} ,
\label{eq:cost_regularization}
\end{equation}
where $\|\cdot\|_2$ denotes the $\ell_2$ norm, $\|\cdot\|_{\mathrm{TV}}$ the standard isotropic TV, 
which enforces smoothness while preserving edges, and $\|\cdot\|_{\mathrm{cTV}}$ the wrapped-phase TV, 
a circular TV adapted for wrapped phase map~\cite{Kamilov:15, weinmann2014total, storath2016exact}. The function of interest $A$ is generally complex-valued in FPT, and thus TV regularization is applied separately to its amplitude $|A|$ and phase $\arg(A)$, with independent weights $\beta_{\mathrm{ampl}}$ and $\beta_{\mathrm{arg}}$ controlling their relative contributions. Additional details of the wrapped-phase TV are provided in Sec.~\ref{sec:TV}.

The reconstruction problem is formulated as
\begin{equation}
\{O^*,P^*\} = \arg\min_{O,P}\, C(O,P),
\end{equation}
and is solved iteratively with an alternating minimization scheme with a plug-and-play (PnP) prior–based update, adapted from~\cite{wang2023fpt}, as summarized in Algorithm~\ref{alg:FPT}.
The recovered object field $O$ provides the wrapped phase map, the primary output of single-wavelength FPT algorithm, 
while the simultaneous estimation of the pupil function $P$ enables digital refocusing, as described in Sec.~\ref{sec:refocus}.

\subsubsection{Wrapped-Phase TV}
\label{sec:TV}
\begin{figure}[b]
\centering
\includegraphics[width=\columnwidth]{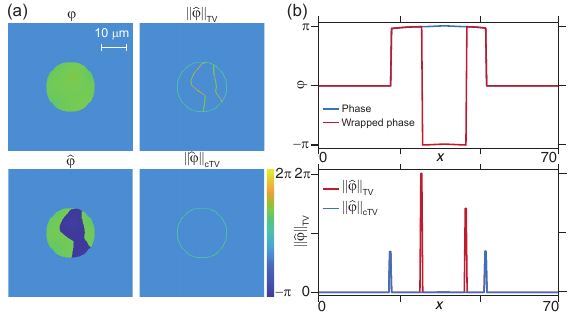}
\caption{Comparison of conventional TV and cTV on wrapped phase data. 
(a)~True phase $\varphi$ (top left) and wrapped phase $\hat\varphi$ (bottom left), together with their gradient magnitudes computed by conventional TV (top right) and cTV (bottom right). 
While conventional TV produces strong artificial edges at the $\pm\pi$ wrapping boundaries, cTV correctly interprets them as continuous on $S^1$ and suppresses such spurious gradients. 
(b) 1D cross sections: the wrapped phase (red) exhibits jumps at $\pm\pi$ (top), which lead to sharp peaks in the conventional TV profile (bottom, red). 
In contrast, cTV (blue) only responds to the true object boundary, yielding physically consistent edge peaks under the $2\pi$ periodicity.}
  \label{fig:TVComparison}
\end{figure}

In phase retrieval algorithms like single-wavelength FPT, the recovered phase is inherently wrapped into $(-\pi,\pi]$. 
This wrapping can cause numerical instability near the boundary: 
two neighboring pixels may correspond to almost identical heights, 
yet be reconstructed as $\pi-\delta_1$ and $-\pi+\delta_2$ due to noise or perturbations. 
Although their true height difference is negligible, 
a naive finite difference of the wrapped phases would erroneously indicate a $2\pi$ jump, 
as illustrated in Fig.~\ref{fig:TVComparison}.

The conventional isotropic total variation (TV) is defined as
\begin{equation}
\|\varphi\|_{\mathrm{TV}}
= \sum_{i,j} \sqrt{ (\Delta_x \varphi_{i,j})^2 + (\Delta_y \varphi_{i,j})^2 }
= \|\mathbf{D}\varphi\|_{1,2},
\end{equation}
where $\Delta_x$ and $\Delta_y$ denote horizontal and vertical forward difference operators, 
 $\mathbf{D}=[\Delta_x,\Delta_y]$, and $\|\mathbf{D}\varphi\|_{1,2}$ the mixed $\ell_{1,2}$ norm.
This formulation assumes access to the unwrapped phase $\varphi$ and therefore cannot be applied directly. 
If applied to the wrapped phase $\hat\varphi$ in the same form, 
it introduces spurious discontinuities when neighboring phases fall across the $\pm\pi$ boundary.  

To respect the $2\pi$-periodicity of phase, we adopt a 
\emph{circular} total variation (cTV)~\cite{weinmann2014total,storath2016exact} that measures first-order differences on the unit circle $S^1$ rather than on the real line. 
Let $\hat\varphi\in(-\pi,\pi]$ denote the wrapped phase. 
For two neighboring pixels, the wrapping operator $\mathcal{W}(\cdot)$, which computes the circular (wrapped) difference, is defined as
\begin{equation}
\mathcal{W}(\Delta\hat\varphi)
= \operatorname{atan2}\!\big(\sin(\Delta\hat\varphi),\,\cos(\Delta\hat\varphi)\big)
\in(-\pi,\pi],
\label{eq:circ-diff}
\end{equation}
which returns the shortest angular displacement on $S^1$. 
The isotropic  cTV is then given by
\begin{align}
\|\hat{\varphi}\|_{\mathrm{cTV}}
&= \sum_{i,j}
\sqrt{
\big(\mathcal{W}(\Delta_x \hat{\varphi}_{i,j})\big)^2 +
\big(\mathcal{W}(\Delta_y \hat{\varphi}_{i,j})\big)^2
} \notag \\
&= \|\mathcal{W}(\mathbf{D}\hat{\varphi})\|_{1,2}.
\label{eq:circ-tv}
\end{align}
By applying $\mathcal{W}(\cdot)$ 
before computing the discrete gradient, cTV penalizes angular variations along the geodesic distance on the circle, 
thus preventing spurious discontinuities at the wrapping boundary 
and yielding smooth, $2\pi$-periodic phase reconstructions.

As illustrated in Fig.~\ref{fig:TVComparison}, the advantage of cTV becomes evident. 
In panel~(a), the true phase $\varphi$ and its wrapped counterpart $\hat\varphi$ are shown together with the corresponding TV responses. 
When applying the conventional TV to $\hat\varphi$, strong artificial gradients appear along the $\pm\pi$ wrapping boundaries, which do not correspond to actual discontinuities in the underlying phase. 
In contrast, cTV correctly interprets these boundary crossings as continuous on $S^1$, and therefore suppresses such spurious edges. 
Panel~(b) further highlights this effect in one-dimensional cross sections: 
while the wrapped phase (red) exhibits jumps at $\pm\pi$, leading to sharp peaks in the standard TV profile, the cTV (blue) remains flat except at the true object boundary. 
This demonstrates that cTV faithfully enforces smoothness consistent with the $2\pi$-periodicity and avoids introducing non-physical artifacts at wrap boundaries.

\subsubsection{Digital Refocusing}
\label{sec:refocus}

In dual-wavelength FPT, reconstructions at different illumination wavelengths suffer from axial focal-plane mismatch due to chromatic aberrations of the imaging system. This misalignment degrades spatial consistency across wavelengths.
We correct this mismatch using digital refocusing via the angular spectrum method~\cite{zheng2013fpm}. Given an in-focus field $U_0(\mathbf{r})$ with wavelength $\lambda_i$, the field can be numerically propagated by an axial distance $z_{\lambda_i}$:
\begin{equation}
U_{z_{\lambda_i}}(\mathbf{r};\lambda_i) = \mathscr{F}^{-1}\!\left\{ \mathscr{F}[U_0(\mathbf{r};\lambda_i)] \,
H_{z_{\lambda_i}}(\mathbf{k};\lambda_i) \right\},
\label{eq:angular_spectrum}
\end{equation}
where 
$H_{z_{\lambda_i}}(\mathbf{k};\lambda_i) 
= \exp\!\left[2\pi j {z_{\lambda_i}} \sqrt{1/\lambda_i^2-|\mathbf{k}|^2}\,\right]$.

In our FPT implementation, digital refocusing is applied prior to the iterative reconstruction. 
During initialization, defocus distances $z_{\lambda_1}$ and $z_{\lambda_2}$ 
are assigned to align the reconstructions at the two wavelengths to a common focal plane. 
These distances can be measured empirically by pre-calibration. 
The corresponding defocus-compensated pupil functions
\begin{equation}
    P'_{\lambda_i}(\mathbf{k}) = P(\mathbf{k})\, H_{z_{\lambda_i}}(\mathbf{k};\lambda_i),
\end{equation}
are then used in the forward model:
\begin{equation}
    I_{n,\lambda_i}(\mathbf{r}) 
= \Big|\mathscr{F}^{-1}\!\big\{\mathscr{F}[O_{\lambda_i}(\mathbf{r}) S_{n,\lambda_i}(\mathbf{r})] 
\, P'_{\lambda_i}(\mathbf{k}) \big\}\Big|^2 .
\end{equation}
Subsequently, the iterative optimization refines $P(\mathbf{k})$, 
so that any residual aberrations beyond defocus are  estimated.

\begin{algorithm}[t]
\DontPrintSemicolon
\caption{Single-wavelength FPT Algorithm}
\label{alg:FPT}

\KwInH{Measurements $\{I^m_{n,\lambda}\}_{n=1}^N$,\\ \HKwBreak illuminations $\{S_{n,\lambda}\}$,\\ \HKwBreak initial pupil $P^{(0)}(\mathbf{k})$, defocus $z_\lambda$, \\ \HKwBreak steps $\eta_O,\eta_P$, iters $T$, \\ \HKwBreak $L_2$ weights $\alpha_{O},\alpha_{P}$, \\ \HKwBreak TV weights $\beta_{\mathrm{O,Ampl}},\beta_{\mathrm{P,Ampl}}$,\\ \HKwBreak cTV weights $\beta_{\mathrm{O,arg}},\beta_{\mathrm{P,arg}}$} 
\KwOutH{Object $O_\lambda(\mathbf{r})$, pupil $P_\lambda(\mathbf{k})$}

\KwInit{$P^{(0)} \leftarrow P^{(0)} \cdot H_{z_\lambda}(\mathbf{k};\lambda)$; \quad
$O^{(0)}(\mathbf{r}) \leftarrow A_0(\mathbf{r})\,e^{i\cdot 0}$}

\For{$t \leftarrow 0$ \KwTo $T-1$}{
  \For{$n \leftarrow 1$ \KwTo $N$}{
    $l_n \leftarrow \left|\mathscr{F}^{-1}\!\left\{
     \mathscr{F}\!\big(O^{(t)} S_{n,\lambda}\big) \, P^{(t)}
   \right\}\right| - \sqrt{I^m_{n,\lambda}}$\;
  }
  $L^{(t)} \leftarrow \sum_n \|l_n\|_2^2$\;
  \For{$A \leftarrow \{O,P\}$}{
    $A^{(t+\frac12)} \leftarrow A^{(t)} - \eta_A \big( \nabla_A L + \alpha_{A}\, A^{(t)} \big)$\;
    $A \leftarrow \mathrm{prox}_{\mathrm{TV}}\!\big(\lvert A^{(t+\frac12)}\rvert;\,\beta_{\mathrm{A,ampl}}\big)$\;
    $\varphi_A \leftarrow \mathrm{prox}_{\mathrm{cTV}}\!\big(\arg(A^{(t+\frac12)});\beta_{\mathrm{A,arg}}\big)$\;
    $A^{(t+1)}_\lambda \leftarrow A \, e^{i \varphi_A}$\;
  }
  }
\KwRet{$O^{(T)}, P^{(T)}$}
\end{algorithm}

\begin{algorithm}[t]
\caption{Dual-wavelength phase unwrapping algorithm}
\label{alg:dwFPT}
\KwInH{Wrapped phases $\hat\varphi_1,\hat\varphi_2$, wavelengths $\lambda_1,\lambda_2$,\\
\HKwBreak tolerance $\tau$, TV weight $\lambda_{\mathrm{TV}}$,\\
\HKwBreak soft-bound weight $\rho$}
\KwOutH{Height $H(x,y)$}

\ForEach{$(x,y)$}{
  $h_i(k_i)\leftarrow \frac{\lambda_i}{4\pi}\big(\hat\varphi_i+2\pi k_i\big),\ i=1,2$\\
  $(K_1,K_2)\leftarrow \arg\min\limits_{k_1,k_2\in[K_{\min},K_{\max}]}\big|h_1(k_1)-h_2(k_2)\big|$\\
  $h^* \leftarrow \tfrac{1}{2}\!\big(h_1(K_1)+h_2(K_2)\big)$\\
  $h_{\min}\!\leftarrow\!\min\{h_1(K_1),h_2(K_2)\}$\\
  $h_{\max}\!\leftarrow\!\max\{h_1(K_1),h_2(K_2)\}$\\
}

\textbf{Global refinement:}\\
Solve \\
$$\begin{aligned}\displaystyle H\leftarrow\arg\min_H \frac12\|H-h^*\|_2^2
+ \lambda_{\mathrm{TV}}\|H\|_{\mathrm{TV}}\\
+ \rho\|\max(0,\,h_{\min}\!-\!H,\,H\!-\!h_{\max})\|_1\end{aligned}$$\\
\Return{$H$}
\end{algorithm}

\subsection{Dual-wavelength Phase Unwrapping Algorithm}
\label{sec:unwrap}

After reconstructing the wrapped phase maps $\hat\varphi_1$ and $\hat\varphi_2$ at wavelengths $\lambda_1$ and $\lambda_2$, phase unwrapping is carried out in two stages (see Algorithm~\ref{alg:dwFPT}):
(i) a pixel-wise wrapped-number search that enforces height consistency across the two wavelengths, and
(ii) a global refinement step that improves noise robustness and enforces piecewise-smooth structure across the height map. 

\subsubsection{Step 1: Pixel-wise wrapped-number search}
The wrapped phases reconstructed at $\lambda_1$ and $\lambda_2$ are
\[
\hat\varphi_{1,2} = \mathcal{W}\!\left(\frac{4\pi}{\lambda_{1,2}}\, h\right) 
= \frac{4\pi}{\lambda_{1,2}}\, h - 2\pi k_{1,2},
\]
with $k_{1,2}\in\mathbb{Z}$ the unknown wrapped-numbers.  
From these, two candidate heights are defined as
\[
h_1 = \Big(\frac{\hat\varphi_1}{4\pi}+\frac{k_1}{2}\Big)\lambda_1,
\qquad
h_2 = \Big(\frac{\hat\varphi_2}{4\pi}+\frac{k_2}{2}\Big)\lambda_2 .
\]
In the ideal noiseless case, one would have $h_1=h_2=h$, 
but in practice noise prevents exact equality.  

To resolve this, we perform a pixel-wise search over integer pairs $(k_1,k_2)$ and select the combination that minimizes the candidate height discrepancy:
\begin{equation}
(K_1,K_2)
= \arg\min_{k_1,k_2}
\left|\frac{\hat\varphi_1+2\pi k_1}{4\pi}\lambda_1
- \frac{\hat\varphi_2+2\pi k_2}{4\pi}\lambda_2\right|.
\label{Equ_unwrapping}
\end{equation}

This wrapped-number search enforces consistency between the two wavelength estimates while avoiding the noise amplification of direct phase differencing in Eq.~\eqref{eq:synphase}. In practice, the search interval $[K_{\min},K_{\max}]$ is chosen according to prior knowledge of the sample height range.

The candidate height estimates $h_1(K_1)$ and $h_2(K_2)$ are combined to form an initial estimate 
$h^*=\tfrac{1}{2}\big(h_1(K_1)+h_2(K_2)\big)$, while  
$h_{\min}=\min\{h_1(K_1),\,h_2(K_2)\}$ and 
$h_{\max}=\max\{h_1(K_1),\,h_2(K_2)\}$ 
define the feasible height bounds for the refinement step.

\subsubsection{Step 2: Global refinement}
To improve accuracy and enforce spatial consistency, the height map $H$ is refined by solving
\begin{equation}
H=\arg\min_{H}\;\frac{1}{2}\|H-h^*\|_2^2
+\lambda_{\mathrm{TV}}\|H\|_{\mathrm{TV}}
+\rho\,\ell_C(H),
\label{Equ_Unwrapping_TV}
\end{equation}
where $\|H\|_{\mathrm{TV}}$ promotes spatial smoothness while preserving discontinuities, and $\ell_C(H)$ is a soft constraint enforcing the feasible height bounds $[h_{\min},h_{\max}]$:
\begin{equation}
\ell_C(H)=\|\max(0,\,h_{\min}-H,\,H-h_{\max})\|_1.
\end{equation}

In Eq.~\eqref{Equ_Unwrapping_TV}, the data term keeps $H$ close to the pixel-wise dual-wavelength estimate $h^*$, the TV term suppresses noise while preserving true step edges, and the soft bound constraint penalizes values outside $[h_{\min},h_{\max}]$ without imposing hard clipping. This refinement reduces wrap-induced outliers and enforces spatial consistency across the surface while maintaining accurate height transitions.

\section{Results}

\subsection{Numerical Simulations of Dual-Wavelength FPT}
\label{sec:Simulation}
The effectiveness of the dual-wavelength FPT approach was validated using full-wave numerical simulations. We employed a Modified Born Series (MBS) solver~\cite{li2025rDTreflect} to generate synthetic intensity measurements for a reflective system with objective NA = 0.28. The illumination was composed of two monochromatic plane waves at $630~\text{nm}$ and $515~\text{nm}$. For each wavelength, the sample was sequentially illuminated from 25 angles, covering an illumination NA up to $0.28$.

\begin{figure*}[t]
\vspace{0.1cm}\includegraphics[width=0.95\textwidth]{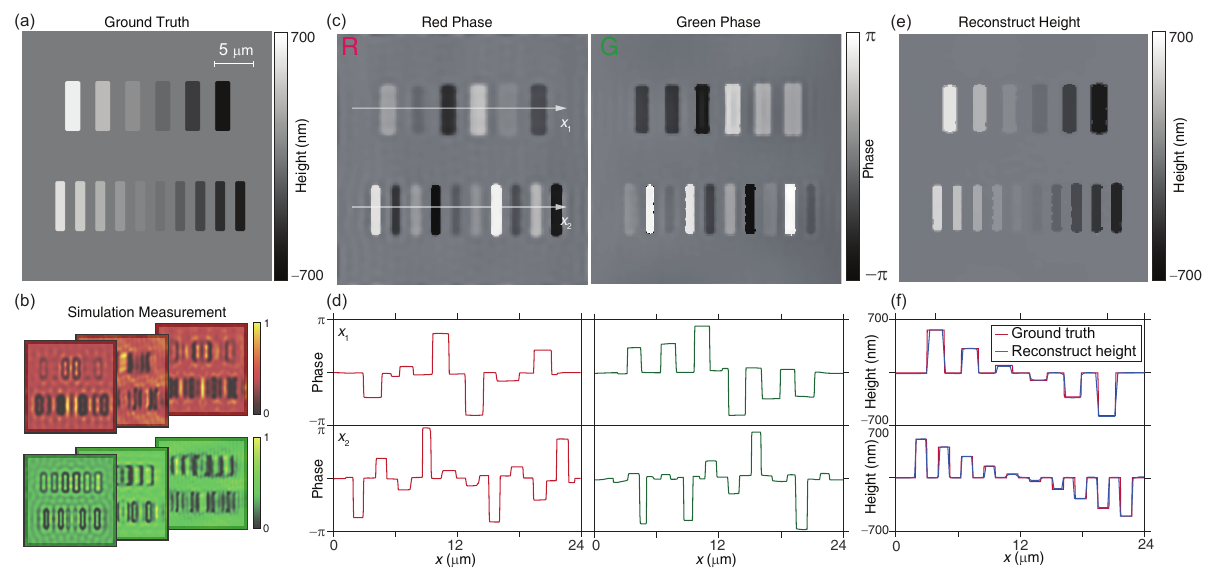}
  \caption{Dual-wavelength FPT reconstruction simulation results.
(a) The ground truth height map of the test sample is composed of structured silicon features with varying topography.
(b) Simulated intensity images generated using the MBS model under red and green illumination.
(c) Wrapped phase reconstructions from single-wavelength FPT under red and green LEDs show phase aliased due to changes in height.
(d) Phase cross-section from single-wavelength FPT under red and green LEDs.
(e) Reconstructed height map using the dual-wavelength phase unwrapping algorithm, which accurately recovers continuous surface profiles.
(f) Height cross-section comparison between the ground truth and the reconstruction demonstrating high quantitative agreement.}
  \label{fig:Simu}
\end{figure*}

The test object is a silicon height map that contains both concave and convex surface profiles as shown in Fig.~\ref{fig:Simu}(a). It comprises two regions: (i) a $1~\text{\textmu m}$-wide strip containing six evenly spaced features with heights from $-0.592$ to $0.592~\text{\textmu m}$, and (ii) a $2~\text{\textmu m}$-wide strip containing ten features with heights from $-0.515$ to $0.515~\text{\textmu m}$. Thus the total height range spans approximately $\pm 0.6~\text{\textmu m}$.

Figures~\ref{fig:Simu}(c,d) show the reconstructions obtained when only a single wavelength is used. As expected, each single-wavelength FPT reconstruction is fundamentally limited to an unambiguous height range of $\lambda/2$: about $316~\text{nm}$ for 630\,nm and $258~\text{nm}$ for 515\,nm. Features exceeding these ranges undergo phase wrapping, causing their apparent heights to alias into the $(0,\lambda/2]$ interval. As a result, they cannot be correctly recovered, appearing as flattened or misestimated steps in the single-wavelength reconstructions.

In contrast, the proposed dual-wavelength phase-unwrapping framework fuses the red and green phase maps to recover a common height profile [Fig.~\ref{fig:Simu}(e,f)]. The reconstructed heights agree well with the ground truth across the full $\pm 0.6~\text{\textmu m}$ range for both narrow and wide features, significantly beyond the $\lambda/2$ limit of either wavelength alone. These simulations confirm that dual-wavelength FPT extends the usable height dynamic range while maintaining the lateral resolution dictated by the synthetic NA.
In this simulation, we consider a symmetric height range \(h \in [-\lambda_s/4,\, \lambda_s/4]\) within the synthetic-wavelength interval \(\lambda_s/2\), and use this prior bound to restrict the wrapped-number search in Eq.~\eqref{Equ_unwrapping} to \(k_i \in [-3,3]\). This setup tests the dual-wavelength unwrapping method in a bidirectional regime around the reference plane.The simulated structure in Fig.~\ref{fig:Simu} spans a height range comparable to that used in our later AR study while maintaining conservative ARs (typically \(\mathrm{AR} < 0.515\)). It serves as a baseline, well-behaved case under the same optical configuration. More aggressive high-AR regimes and their impact on reconstruction fidelity are examined in Sec.~\ref{sec:Discussion}.

\subsection{Benchmark of Proposed Phase Unwrapping Method}

\begin{figure*}[t]
\vspace{0.1cm}\includegraphics[width=0.95\textwidth]{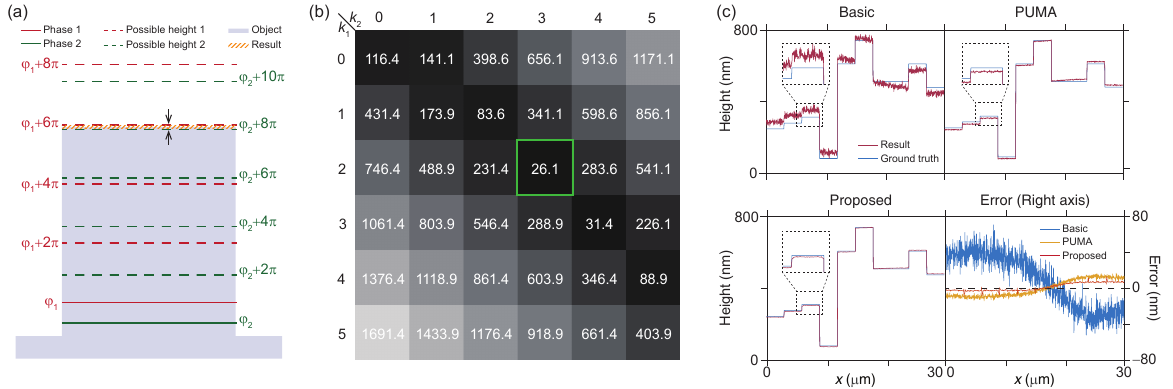}
\caption{Dual-wavelength phase-unwrapping mechanism and performance comparison. 
(a) Schematic of the wrapped-number search. Each wrapped phase $\varphi_1$ and $\varphi_2$ maps to multiple candidate heights differing by integer multiples of $2\pi$. Cross-wavelength consistency is used to identify the correct physical height (orange line). 
(b) Error matrix over integer-pair combinations $(k_1,k_2)$. The optimal pair (green box) is the global minimum of the consistency metric, yielding the most reliable height estimate. 
(c) Quantitative comparison of three unwrapping strategies: the baseline ``Basic'' phase difference method, PUMA, and the proposed approach. The baseline method amplifies noise and deviates significantly from the ground truth. PUMA improves accuracy but is computationally expensive and retains larger residual errors. The proposed method closely matches the ground truth across all regions with substantially lower noise sensitivity.}
\label{fig:Unwrap}
\end{figure*}

To benchmark the proposed dual-wavelength phase unwrapping method, we conduct a numerical test case and compare its performance with several alternative algorithms. The setup and results are summarized in Fig.~\ref{fig:Unwrap}.

The test object is a piecewise-constant height profile with multiple steps of different heights, most of which exceed the single-wavelength $\lambda/2$ range. 
For each wavelength, we add a slowly varying background phase term to mimic low-frequency aberrations, and then superimpose small pixel-wise Gaussian noise. Together these two components emulate realistic phase errors in single-wavelength FPT reconstructions.

Figure~\ref{fig:Unwrap}(a) illustrates how the two wrapped phases $\hat\varphi_1$ and $\hat\varphi_2$ each map to a family of candidate heights that differ by integer multiples of $2\pi$. Each horizontal dashed line denotes a distinct wrapping number $k_1$ or $k_2$ and therefore a different absolute height. Even when the wrapped phases are noisy, Eq.~\eqref{Equ_unwrapping} evaluates all integer pairs $(k_1,k_2)$ and identifies the pair $(K_1,K_2)$ whose corresponding heights are most consistent across wavelengths. This wrapped-number search recovers the correct physical height (orange curve) without forming a noisy synthetic phase through direct differencing.
Figure~\ref{fig:Unwrap}(b) visualizes the $(k_1,k_2)$ optimization as an error matrix over all integer pairs. Each entry corresponds to the error metric in Eq.~\eqref{Equ_unwrapping}, and the optimal pair (green box) is the global minimum of this matrix.

Figure~\ref{fig:Unwrap}(c) compares three phase-unwrapping algorithms under identical background bias and noise conditions. 
The Basic method forms the synthetic phase $\varphi_s$ (Eq.~\eqref{eq:synphase}) and unwraps it directly; as expected, the differencing step amplifies noise and produces substantial deviations from the ground-truth height profile. PUMA is a state-of-the-art dual-wavelength unwrapping algorithm~\cite{bioucas2007puma}; it substantially reduces reconstruction error but requires solving a large-scale graph-cut problem and is computationally demanding (in our implementation, PUMA requires \(\sim 40\)~s for this data). In contrast, the proposed method Sec.~\ref{sec:unwrap} closely matches the ground truth across all regions, exhibits far lower noise sensitivity than the Basic method, and achieves this performance at a fraction of PUMA’s computational cost (about $2$~s on the same data). This simulation demonstrates that the proposed approach is both noise robust and computationally efficient.

\subsection{Experimental Setup}

The FPT system employs an LED array with 25 illumination angles arranged on three concentric rings (Fig.~\ref{fig:Pipeline}(a)). Each LED package (Kingbright APTF1616SEEZGQBDC) provides red, green, and blue emission with central wavelengths of 630\,nm, 515\,nm, and 460\,nm~\cite{wang2023fpt}. The array is placed at the front focal plane of the first lens in a 4$f$ relay, which maps each LED position to the back focal plane of the objective. 

The array consists of a central LED surrounded by two concentric rings. The physical radii of the inner and outer rings are 4.2~mm and 8.4~mm, with 8 and 16 LEDs on each ring, respectively. A pair of relay lenses (Thorlabs ACT508-250-A and AC508-150-A) and a cube beam splitter (Thorlabs, CCM1-BS013) are chosen so that the outer bright-field ring is mapped to the edge of the objective pupil aperture. Under this configuration, the corresponding illumination numerical apertures of the two rings are approximately 0.14 and 0.28.

The LED board is driven by TLC5955 current drivers (Texas Instruments). We time-multiplex both angle and color: for each measurement sequence, red and green LEDs are sequentially activated at each angle, providing the dual-wavelength dataset required by our algorithm.

Samples are mounted on a multi-axis translation stage (Newport 37 series stages, Thorlabs NanoMax300 with APY002 actuators), which allows fine adjustment of focus and lateral positioning. The reflected intensity images are recorded by a monochrome CMOS camera (Imaging Source, DMK38UX541) with a pixel size of 2.74~\textmu m and  $4504\times4504$ pixels. During acquisition, LEDs are switched on one at a time, and the camera is triggered synchronously. The typical exposure time is 500~ms per frame, leading to a total acquisition time on the order of a few minutes for a full dual-wavelength, multi-angle dataset.

The field of view (FOV) is limited by the camera sensor and is approximately $1.2\times1.2~\text{mm}^2$ in the object plane. With an objective NA of 0.28 and a maximum illumination NA of 0.28, the effective synthetic NA for FPT is $\mathrm{NA}_{\mathrm{syn}} = \mathrm{NA}_{\mathrm{obj}} + \mathrm{NA}_{\mathrm{illum}} = 0.56$, which sets the achievable lateral resolution for both wavelengths used in the reconstructions.

\subsection{Single-Wavelength FPT Validation}

\begin{figure}[h]
    \centering
\includegraphics[width=0.48\textwidth]{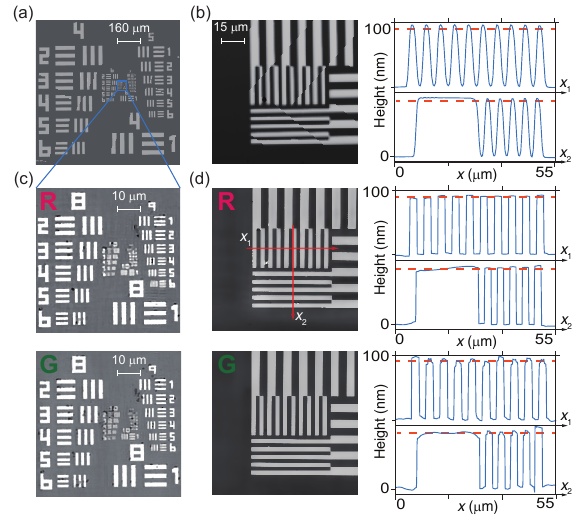}
  \caption{Experimental validation of single-wavelength FPT.
(a) Bright-field reflectance image of a USAF resolution target used to assess amplitude resolution. 
(b) Ground-truth height profile of a calibrated 100\,nm step structure measured by a Zygo interferometer. 
(c) Reconstructed amplitude images under red and green LED illumination, demonstrating recovery of high-frequency features. 
(d) Reconstructed phase maps for the red and green wavelengths, with corresponding line profiles showing agreement with the Zygo reference. Red dotted lines: expected height of the structures.}
  \label{fig:Exper1}
\end{figure}

Before evaluating the dual-wavelength phase unwrapping procedure, we first assess the performance of the single-wavelength reconstruction (Algorithm~\ref{alg:FPT}) at individual wavelengths. Two standard samples were used: a USAF 1951 resolution target to quantify lateral resolution  and a calibrated 100\,nm phase step to evaluate height accuracy.

Reconstruction was performed on an $80\times 80~\mu\text{m}^2$ region.
Under red illumination, the system resolves Group~9, Element~6 of the USAF target, corresponding to a theoretical resolution of $0.56~\mu\text{m}$.
Under green illumination, the system resolves Group~10, Element~1, corresponding to a theoretical resolution of $0.46~\mu\text{m}$ (Fig.~\ref{fig:Exper1}(c)). 
These measured resolutions agree well with the diffraction-limited performance of the system, ($\lambda_i/\mathrm{NA}_{\mathrm{syn}}$), which is $0.56~\mu\text{m}$ at 630\,nm and $0.46~\mu\text{m}$ at 515\,nm. 

Phase accuracy was evaluated using the 100\,nm step-height sample (Fig.~\ref{fig:Exper1}(b)). 
The reconstructed wrapped phase reproduces the expected height profile at both wavelengths, indicating that the single-wavelength algorithm yields quantitatively accurate phase within its unambiguous range. 
Since the step height lies within the single-wavelength unambiguous range, phase unwrapping was not required. These results confirm that single-wavelength FPT algorithm produces accurate wrapped phase maps that can be reliably used as inputs to the subsequent dual-wavelength fusion.

\subsection{Digital Refocusing in FPT}

\begin{figure}[h]
  \centering
  \includegraphics[width=0.48\textwidth]{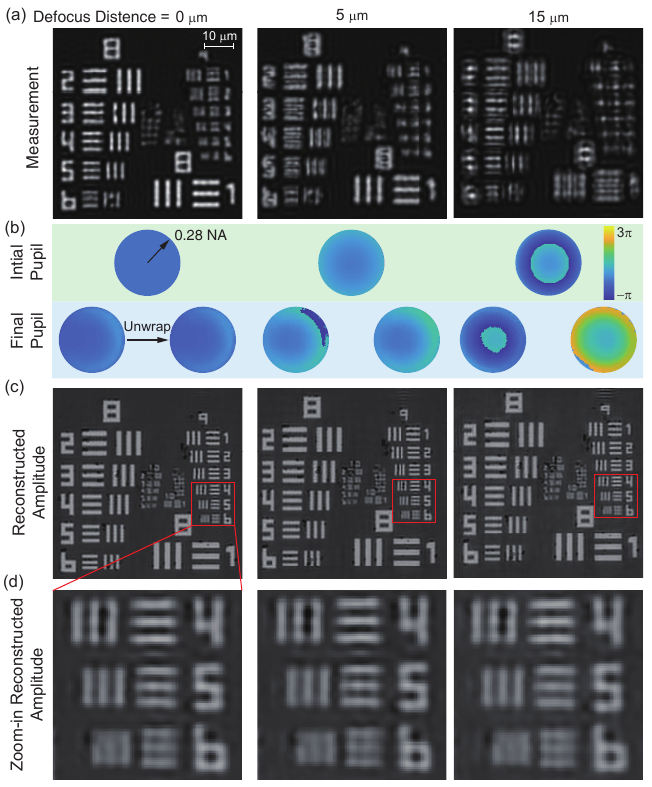}
\caption{Demonstration of digital refocus in FPT. 
(a) Raw intensity measurements acquired at axial offsets 
$z=\{0,\,5,\,15\}\,\mu\text{m}$ from best focus. 
(b) Pupil phase before (top) and after (bottom) reconstruction with pupil-phase unwrapping.  
(c) Reconstructed object amplitudes using the digitally refocused pupil; red boxes indicate regions of interest. 
(d) Magnified views of the marked regions show that image sharpness and feature contrast are preserved across defocus levels, demonstrating that digital refocusing combined with pupil-phase estimation yields robust, defocus-invariant reconstruction.}
  \label{fig:Refocus}
\end{figure}

As discussed earlier, dual-wavelength FPT is affected by chromatic defocus. In our system, calibration showed that the red and green focal planes differ by approximately $10~\mu\text{m}$. To avoid mechanical refocusing, all measurements were acquired at the midpoint between the two foci ($5~\mu\text{m}$ from each best-focus plane), and digital refocusing was applied independently to each wavelength.

To evaluate this strategy, we acquired measurements of a USAF 1951 target over intentional defocus offsets from $5$ to $15~\textmu\text{m}$. The corresponding defocus-compensated pupil functions were computed and used as initialization for reconstruction.

Figure~\ref{fig:Refocus} shows representative reconstructed object amplitudes and pupil phases. Despite the visible defocus artifacts in the raw data, digital refocusing produces consistent, high-fidelity reconstructions. The recovered pupil phases remain smooth and free of discontinuities, indicating that the angular-spectrum initialization preserves the underlying pupil structure. These results confirm that digital refocusing effectively compensates chromatic defocus across axial offsets up to $15~\mu\text{m}$, enabling accurate dual-wavelength imaging without mechanical focal adjustment.

\subsection{Dual-wavelength FPT Validation in Experiments}
\label{sec:Experimenet}
\begin{figure*}[htpb]
    \centering
\vspace{0.1cm}\includegraphics[width=0.95\textwidth]{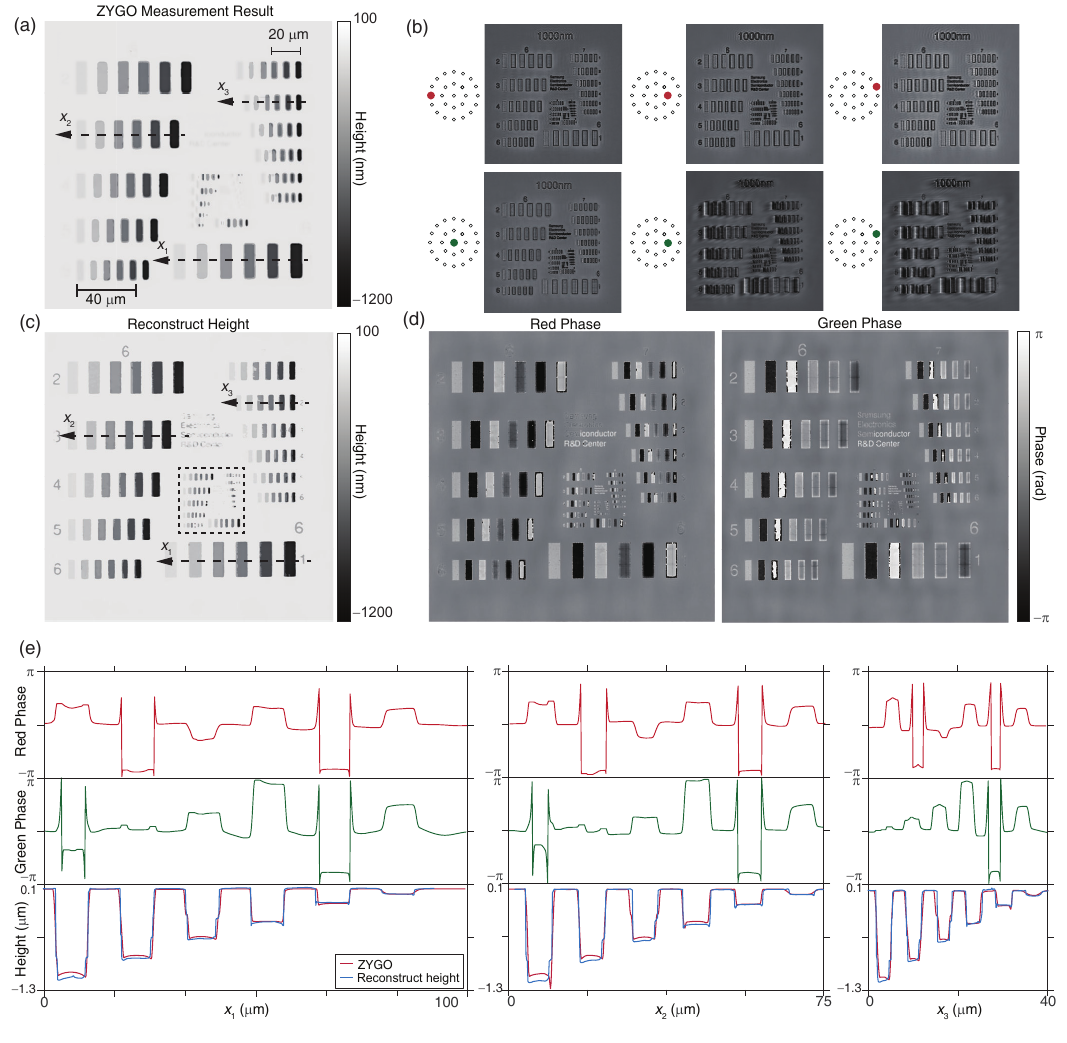}
\caption{Experimental validation of dual-wavelength FPT on complex topographies.
(a) Reference height map of a patterned silicon surface with multiple step-height regions, measured by a Zygo interferometer. 
(b) Representative intensity measurements under red and green LED illumination across different angles. 
(c) Reconstructed height map using the proposed dual-wavelength FPT, accurately recovering both low- and high-relief features; the boxed region denotes the small-scale area further examined in Fig.~\ref{fig:Exper2_limit}. 
(d) Single-wavelength reconstructed phase maps. 
(e) Cross-sectional phase and height profiles from (c) and (d). The red and green curve in the first and second row show the phase reconstructed under two wavelength LED illumination. The dual-wavelength reconstruction (blue) closely matches the Zygo reference (red, third row) across all regions.}
  \label{fig:Exper2}
\end{figure*}

We further validated the dual-wavelength FPT on fabricated samples with a broad range of step heights. The test structure [Fig.~\ref{fig:Exper2}(a)] consists of a silicon substrate with a thermally grown $\mathrm{SiO}_2$ layer, patterned into multiple groups of rectangular trenches with different widths and depths. Each group contains six nominally identical features with heights spanning from $80$~nm up to $1000$~nm, while the lateral widths vary from $7.88$~\textmu m down to $0.55$~\textmu m. This design provides a wide range of ARs and lateral scales within a single FOV.

Representative intensity measurements under red and green LED illumination at multiple angles are shown in Fig.~\ref{fig:Exper2}(b), forming the dataset used for FPT reconstruction. After applying the single-wavelength FPT independently at each color, the reconstructed phase maps [Fig.~\ref{fig:Exper2}(d)] exhibit clear phase wrapping when the local height exceeds the $\lambda/2$ limit of the corresponding wavelength. In these regions, the recovered phase saturates and the step heights alias, preventing a unique physical height assignment from a single wavelength.

Using the proposed dual-wavelength unwrapping algorithm, together with digital refocusing to correct chromatic defocus, we obtain a continuous height map across the entire structure (Fig.~\ref{fig:Exper2}(c)). Cross-sectional phase and height profiles are shown in Fig.~\ref{fig:Exper2}(e). The dashed red curve in the first row shows the phase profile without digital refocusing, which is noticeably distorted. After applying digital refocus and dual-wavelength unwrapping, the reconstructed height profiles (blue) closely match the Zygo reference measurements (red) in both low- and high-relief regions. These results demonstrate that dual-wavelength FPT accurately recovers $\sim 1~\mu\text{m}$ height variations at sub-micron lateral resolution, a regime in which single-wavelength FPT exhibits severe phase-wrapping artifacts.

Importantly, comparison between Fig.~\ref{fig:Exper2}(a) and (c) highlights an additional advantage: although Zygo interferometry serves as the reference standard in this study, it does not resolve the finest lateral features and fails to measure the text patterns present in the sample. In contrast, FPT maintains sub-micron lateral resolution even for structures with large height variations, enabling  characterization of features that lie beyond the lateral-resolution limit of the Zygo system.

\section{Discussion}
\label{sec:Discussion}

FPT enables large FOV, high lateral resolution, and quantitative topography reconstruction~\cite{wang2023fpt}. However, its forward model assumes that the object is a surface-like (quasi–2D) structure that remains illuminated and visible under all incident angles.

For the micron-scale features and height ranges considered in our full-wave simulations (Sec.~III-A, Fig.~\ref{fig:Simu}) and in the larger trenches of the experimental sample in Fig.~\ref{fig:Exper2}(c) (outside the boxed region), this approximation, together with the physical NA limits of the system, is well satisfied. In the simulations, the deepest structures reach heights of about $0.515~\mu\mathrm{m}$ for a $1~\mu\mathrm{m}$ opening ($\mathrm{AR}\approx 0.52$), and in the outer experimental trenches the depths are on the order of $1~\mu\mathrm{m}$ for an opening of approximately $2.19~\mu\mathrm{m}$ ($\mathrm{AR}\approx 0.46$); in this regime the dual-wavelength FPT reconstructions closely follow the ground truth with only small residual errors. In contrast, the high-AR structures highlighted by the boxed region in Fig.~\ref{fig:Exper2}(c) probe a more extreme regime, where comparable heights are combined with much smaller openings, driving the aspect ratio beyond $\mathrm{AR}\approx 0.5$; in this region, the reconstructed height maps begin to show loss of contrast and systematic deviations, indicating that the underlying assumptions of the surface-based forward model are being violated.

When the object height becomes comparable to its lateral feature size, geometric occlusion and multiple reflection paths violate this assumption. As a result, structures with high lateral frequencies and pronounced height variations exhibit reconstruction artifacts that cannot be mitigated simply by increasing the illumination NA. In the following, we quantify how these physical constraints and forward-model approximations jointly limit the accuracy of single-wavelength FPT, identify the primary limitations of the technique, and introduce performance metrics through a series of numerical studies. We also provide a detailed error analysis of our phase unwrapping algorithm.

\subsection{Quantities of Interest for FPT Performance}

Complex topographic surfaces impose coupled geometric and wave-optical constraints on FPT.  
To characterize these effects in a manner directly relevant to our test samples in Fig.~\ref{fig:Exper2}, which contain deep, sub-micron trenches recessed beneath a top surface, we introduce two quantities of interest:

\begin{figure*}[htbp]
  \centering
  \includegraphics[width=0.95\textwidth]{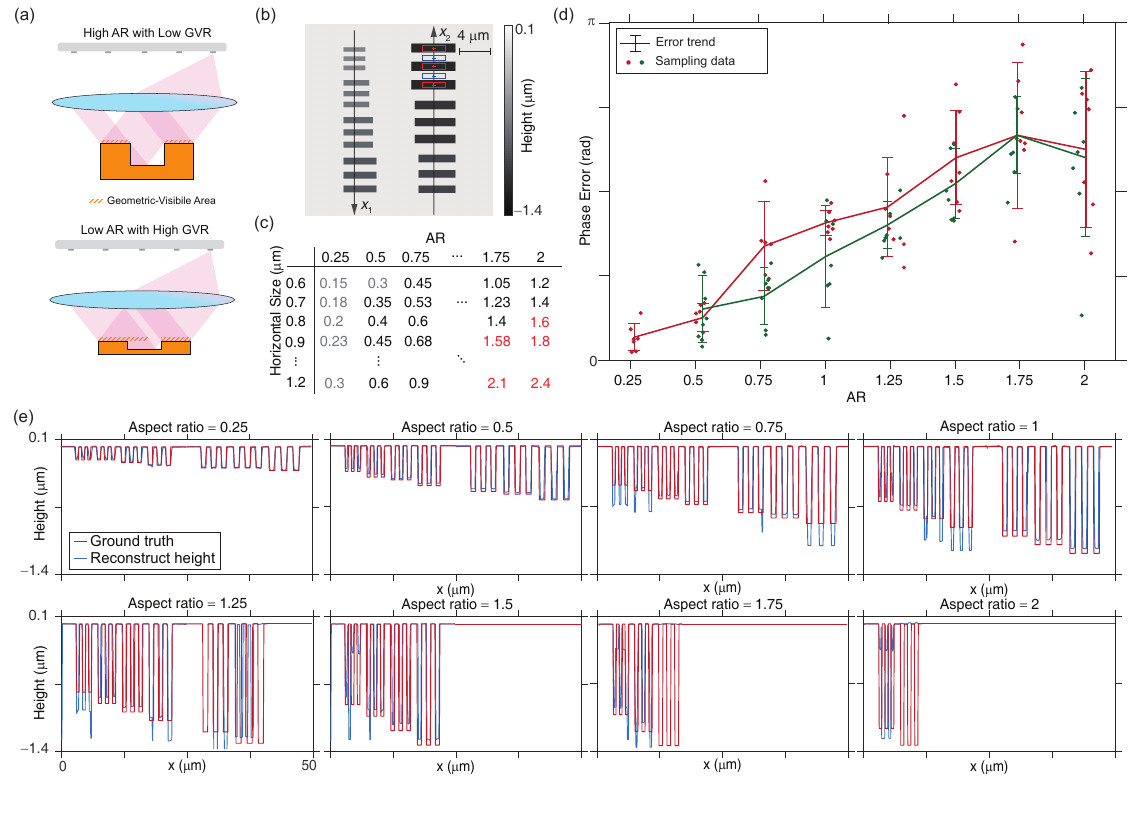}
    \caption{AR dependence of phase error and its impact on dual-wavelength reconstruction.
    (a)Illustration of self-occlusion under high-NA illumination. Deep, narrow features lose direct illumination at large angles, reducing the GVR. Orange hatching marks the region that remains directly illuminated.  
    (b) Layout of test patterns. Colored boxes mark regions used for statistics at different ARs. The red and blue areas are used to delineate the object's height and background.
    (c) Design matrix of opening widths $W$ and AR values used in this study. Our structures are micro-scale with $\leq0.6~\mu\mathrm{m}\leq W\leq1.2~\mu\mathrm{m}$. To enrich statistics, we additionally include many features whose heights lie outside the dual-wavelength unambiguous range. Gray data points indicate heights that do not require phase unwrapping, while red points denote those whose heights lie outside the dual-wavelength unambiguous range.
    (d) Absolute phase error $\Delta\varphi=|\varphi_{\text{measure}}-\varphi_{\text{ground truth}}|$ versus AR. Solid lines indicate the error trend. Dots denote the data reconstructed from the sample.
    (e) Dual-wavelength unwrapped height varies according to changes in AR: the reconstruction error remains small for $\mathrm{AR}\le 0.5$, while for $\mathrm{AR}\ge 0.75$ large deviations appear due to incorrect selection of the wrapped-number pair in the unwrapping step.}
    \label{fig:AR}
\end{figure*}

\textbf{1) Aspect ratio (AR).}
A geometric descriptor,
\[
\mathrm{AR}=\frac{H}{W},
\]
where \(H\) is the feature depth and \(W\) is the lateral opening.  
High-AR structures are deep relative to their width and are therefore more susceptible to shadowing and degradation of phase transfer.  
This metric is directly relevant to the narrow trench features in Fig.~\ref{fig:Exper2}, where depth-to-width geometry strongly impacts reconstruction fidelity.

For intuition, we also consider a simple geometric-visibility ratio (GVR) that estimates the fraction of a recessed bottom surface that remains directly illuminated at the most oblique illumination angle.  
Let \(W_{\mathrm{vis}}\) be the portion of the bottom not shadowed by the sidewalls in a ray-optics model, and define
$\mathrm{GVR}=\frac{W_{\mathrm{vis}}}{W}$.
Here, \(\mathrm{GVR}=1\) indicates full visibility and \(\mathrm{GVR}=0\) complete shadowing.  
We use GVR only as an intuitive, ray-based proxy: in the sub-micron regime studied here, full-wave simulations and experiments show that diffraction and 3D scattering cause phase-transfer breakdown well before \(\mathrm{GVR}\) reaches zero, making the practical AR limit considerably more restrictive than this simple geometric bound.

\textbf{2) Phase Modulation Transfer Function.}
A frequency-domain measure of reconstructed phase fidelity.  
For a periodic phase object with true modulation $\varphi_{\text{true}}(f)$ at spatial frequency $f$, we define
\begin{equation}
\mathrm{MTF}_\varphi(f)
=\frac{\lvert \hat{\varphi}(f) \rvert}{\lvert \varphi_{\text{true}}(f) \rvert},
\label{eq:phase_mtf}
\end{equation}
where $\hat{\varphi}(f)$ is the recovered phase amplitude.  
Values near 1 indicate faithful phase transfer, while values near 0 indicate loss of that lateral frequency.  
ph-MTF is particularly informative for structures in Fig.~\ref{fig:Exper2}, as it quantifies how varying varying trench widths affect high-frequency phase information.

\subsection{Effect of AR in simulation}

In our system, the maximum oblique illumination angle is set by the illumination NA.  
A simple ray-optics estimate suggests that the directly illuminated portion of a trench bottom decreases roughly as 
\(\mathrm{GVR} \approx 1 - 2\,\mathrm{AR}\cdot\mathrm{NA}_{\mathrm{illum}}\),  
implying full shadowing at \(\mathrm{AR}_{\text{geo}} \approx 1/(2\,\mathrm{NA}_{\mathrm{illum}})\), or about \(1.8\) for \(\mathrm{NA}_{\mathrm{illum}}=0.28\).  
This quantity is used only as a rough, first-order reference: in the sub-micron regime studied here, full-wave simulations and experiments already show substantial phase-transfer degradation at \(\mathrm{AR}\approx 0.5\!-\!0.75\), well before the ray-based visibility reaches zero.  
This indicates that diffraction and 3D scattering impose a far stricter practical AR limit than suggested by the simple geometric-visibility approximation.

Figure~\ref{fig:AR} shows the simulated reconstruction error as a function of AR.  
With \(\mathrm{NA}_{\mathrm{illum}}\approx0.28\), the phase error remains small up to \(\mathrm{AR}\approx 0.5\) and then increases sharply at \(\mathrm{AR}\approx 0.75\).  
Beyond this point, the reconstruction becomes unstable and unwrapping failures occur.  
This behavior is consistent across all simulated feature sizes, including those outside the dual-wavelength unambiguous range.

Although the geometric bound permits ARs up to \(\sim1.8\), the simulations break down at \(\mathrm{AR}\approx 0.75\).  
In the sub-micron regime considered here (\(W \le 1.2~\mu\mathrm{m}\)), this earlier failure is attributed to strong 3D wave-scattering effects that degrade phase-transfer fidelity before full geometric shadowing.

Finally, the AR limit observed in Fig.~\ref{fig:AR} is determined by the single-wavelength FPT forward model, not by the dual-wavelength unwrapping.  
Dual-wavelength fusion extends the unambiguous height range but does not change the physical measurement constraints, leading to a practical usable AR of approximately 0.75 in our configuration.

\subsection{AR effects in experimental reconstructions}

\begin{figure}[t]
  \centering
  \includegraphics[width=0.45\textwidth]{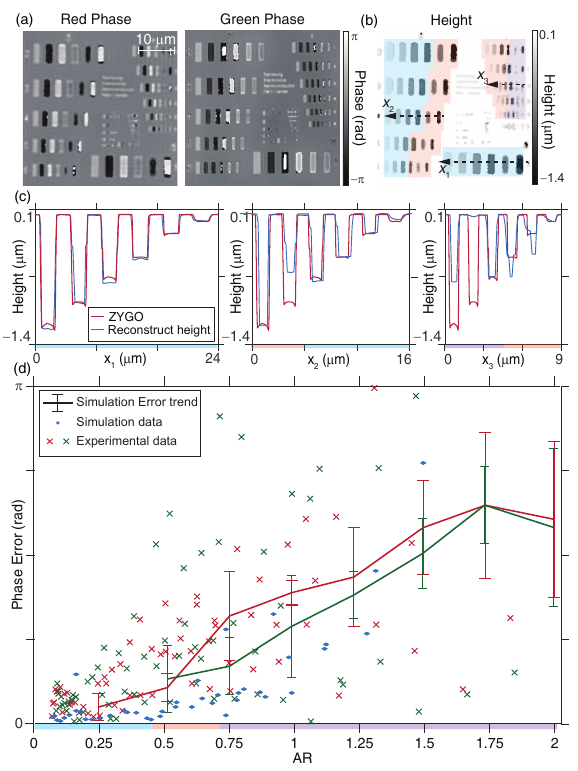}
  \caption{Experimental validation of AR–dependent performance.
    (a) High-resolution single-wavelength FPT reconstructions (wrapped phase) for the region highlighted in Fig.~\ref{fig:Exper2}(c) under red (left) and green (right) illumination.
    (b) Dual-wavelength reconstructed height map of the same region. Cyan, orange, and purple overlays indicate zones with \(\mathrm{AR}<0.5\), \(0.5\le\mathrm{AR}<0.75\), and \(\mathrm{AR}\ge0.75\), respectively; arrows mark the line-scan positions \(x_1\)–\(x_3\).
    (c) Height profiles along \(x_1\)–\(x_3\): dual-wavelength FPT (blue) compared with the Zygo reference (red). Colored bars on the axes denote the corresponding AR regimes.
    (d) Single-wavelength absolute phase error versus AR. Red and green crosses denote the experimental phase errors relative to the Zygo reference for the two illumination wavelengths. Solid curves (with error bars) show the simulated trend from Fig.~\ref{fig:AR}. The blue dots show the simulation phase error data from Fig ~\ref{fig:LimitGeometry} The colored bar along the axis again marks the low-, intermediate-, and high-AR ranges.}
  \label{fig:Exper2_limit}
\end{figure}

The AR-dependent limitations observed in simulation also appear in our experiments.  
Figure~\ref{fig:Exper2_limit} shows a zoomed-in view of the boxed region in Fig.~\ref{fig:Exper2}(c).  
In this inner region, the nominal trench depths are identical to those in the outer pattern, but the lateral openings shrink from roughly \(2~\mu\text{m}\) down to below \(0.5~\mu\text{m}\), approaching and eventually surpassing the system’s lateral-resolution limit. 
Based on the AR inferred from the design (supported by our simulations), we classify features into three groups: cyan regions with \(\mathrm{AR}<0.5\) (reliable), orange regions with \(0.5\le\mathrm{AR}<0.75\) (intermediate), and purple regions with \(\mathrm{AR}\ge0.75\) (unreliable).

Figure~\ref{fig:Exper2_limit}(d) summarizes the AR dependence by plotting the single-wavelength phase error versus AR.  
The experimental data under both wavelengths closely follow the simulated trend, with error magnitude and variance increasing markedly as AR increases.

For trenches in the cyan region (\(\mathrm{AR}<0.5\)), the dual-wavelength FPT reconstructions agree closely with the Zygo reference, with all 24 features having absolute height errors below \(50~\text{nm}\).  
In the intermediate orange region (\(0.5\le\mathrm{AR}<0.75\)), reconstructed contrast begins to degrade and local fluctuations increase; among 27 features, 13 exceed \(100~\text{nm}\) absolute error, indicating occasional wrap-selection errors.  
In the purple region (\(\mathrm{AR}\ge0.75\)), all 15 features show errors above \(100~\text{nm}\), and fine structures are significantly blurred or lost, consistent with the simulated collapse of high-frequency phase transfer and increased unwrap failures in Fig.~\ref{fig:AR}.

Overall, these measurements confirm that, for the NA and wavelength configuration used here, dual-wavelength FPT provides reliable, high-resolution topography for quasi-2D structures with aspect ratios up to roughly \(\mathrm{AR}\approx 0.75\).  
Beyond this regime, geometric visibility and 3D scattering increasingly violate the 2D surface-based forward model assumption.  
The AR- and visibility-based analysis introduced here provides a practical guideline for designing and applying FPT systems in real metrology environments.

\begin{figure}[ht]
  \centering
  \includegraphics[width=0.48\textwidth]{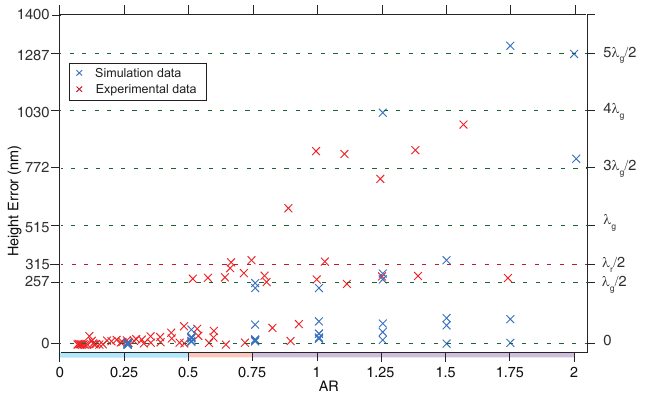}
\caption{Unwrapped height error versus AR.
Absolute height error \( |\Delta h| \) for simulated (blue crosses) and experimental (red crosses) structures as a function of AR. Horizontal dashed lines mark the single-wavelength half-wavelength limits \( \lambda_g/2 \) and \( \lambda_r/2 \). For low AR (\(\mathrm{AR} \le 0.5\)), all points lie close to zero, indicating reliable wrapped-number selection and accurate dual-wavelength unwrapping. As AR increases beyond \(\mathrm{AR} \approx 0.5\), the error distribution broadens and a subset of points clusters near the dashed levels, corresponding to failed unwrapping events where noise and model mismatch lead to an incorrect wrapped-number pair \((K_1, K_2)\) and a resulting height bias of one or more half-wavelengths.}
  \label{fig:HeightError}
\end{figure}

\subsection{Unwrapped height error}

To quantify how aspect ratio affects the final dual-wavelength reconstruction, Fig.~\ref{fig:HeightError} plots the absolute unwrapped height error \( |\Delta h| = |h_{\text{rec}} - h_{\text{GT}}| \) as a function of AR, combining both simulation and experimental data. Blue markers denote simulated structures, and red markers correspond to experimental trenches whose nominal heights fall within the dual-wavelength unambiguous range.

At low aspect ratios (\(\mathrm{AR} \le 0.5\)), both simulation and experiment exhibit small height errors, well below the single-wavelength half-wavelength limits \(\lambda_r/2\) and \(\lambda_g/2\). This indicates that the dual-wavelength wrapped-number search reliably selects the correct integer pair \((K_1, K_2)\) in Eq.~\eqref{Equ_unwrapping}. 

As AR increases toward and beyond \(\mathrm{AR} \approx 0.75\), the error distribution broadens. While many features remain accurate, a subset exhibits large errors clustered around discrete levels near \(\lambda_r/2\) and \(\lambda_g/2\) (horizontal dashed lines). These outliers correspond to failed unwrapping events in which noise or model mismatch leads to an incorrect wrapped-number-pair selection, shifting the recovered height by one or more half-wavelengths.

The emergence of these discrete error bands with increasing AR mirrors the growth in single-wavelength phase error seen in Fig.~\ref{fig:AR}(d). This confirms that failures at high AR arise primarily from degradation of the underlying surface-model phase transfer, due to geometric shadowing and 3D scattering, rather than limitations of the dual-wavelength phase unwrapping algorithm itself. Taken together with the trends in Fig.~\ref{fig:AR}, Fig.~\ref{fig:Exper2_limit}, and Fig.~\ref{fig:HeightError}, these results indicate that, in the micro-scale regime considered here, dual-wavelength FPT provides reliable, quantitatively accurate height reconstructions for quasi-2D structures up to aspect ratios of about $\mathrm{AR}\approx 0.5$, while beyond this regime the breakdown of surface-based phase transfer dominates and dual-wavelength fusion can no longer compensate for these model errors.

\subsection{Ph-MTF Analysis}
To complement the AR- and height-error-based characterization above, we next examine FPT performance from a lateral-frequency perspective using the ph-MTF defined in Eq.~\eqref{eq:phase_mtf}. While AR summarizes geometry at the feature level, ph-MTF provides a complementary, frequency-domain view of how the amplitude of reconstructed phase modulations decays with spatial frequency.

\begin{figure}[t]
  \centering
  \includegraphics[width=0.5\textwidth]{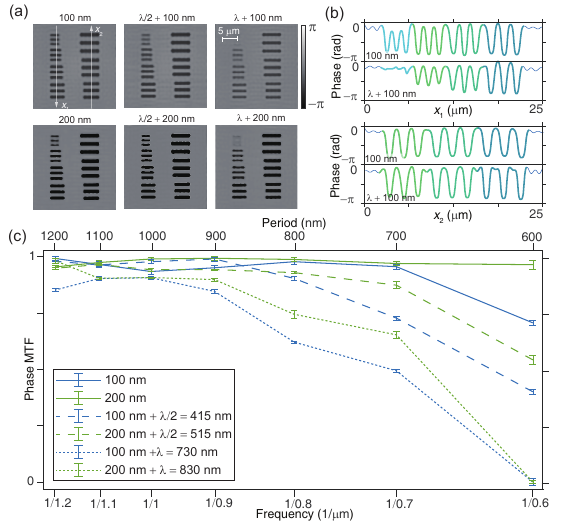}
\caption{
Ph-MTF behavior in FPT.  
(a) Simulated reconstructions of a resolution target for multiple heights; lateral details vanish for the two highest AR structures due to shadowing and multi-path scattering.  
(b) Cross-sectional phase profiles along two line scans ($x_1$, $x_2$) for the 100-nm case and the $\lambda + 100$-nm (i.e. $4\pi$-wrap) case, showing that high-frequency features are not reconstructed at large $H$. 
(c) phMTF versus spatial frequency for step heights $H=100$–$830$ nm. Higher $H$ produces an earlier high-frequency roll-off, consistent with reduced geometric visibility. Error bars denote one standard deviation from five independent reconstructions at SNR = 30 dB.  
}
  \label{fig:LimitGeometry}
\end{figure}

The structures in Fig.~\ref{fig:LimitGeometry}(a) were generated in simulation to isolate the effect of physical height from the effect of phase wrapping.  
To ensure that wrapping alone does not influence the ph-MTF, each base height (100~nm and 200~nm) was also shifted by \(\lambda/2\) and \(\lambda\).  
These height-shifted cases produce identical or nearly identical wrapped phases, so if the 2D phase assumption held, all curves with the same wrapped phase would exhibit identical ph-MTF behavior.

The simulated results show otherwise.  
Even when the wrapped phase is the same, the ph-MTF degrades systematically as the true physical height increases.  
The roll-off is modest for 100~nm features but becomes pronounced for 200~nm and for the height-shifted variants (200~nm+\(\lambda/2\), 200~nm+\(\lambda\)).  
This confirms that the reduction in high-frequency phase transfer is not caused by phase wrapping, but by true height-dependent effects such as partial visibility loss and wavefront distortion in deeper trenches.

\begin{figure}[t!]
  \centering
  \includegraphics[width=0.45\textwidth]{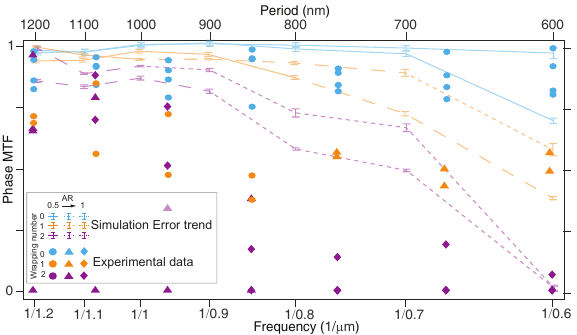}
  \caption{Comparison of simulated and experimental ph-MTF.
Ph-MTF plotted versus lateral spatial frequency (bottom axis) and the corresponding period (top axis). 
Solid and dashed curves with error bars show the simulated ph-MTF from Fig.~\ref{fig:LimitGeometry}(c), grouped by AR ranges and replotted here as reference. 
Markers denote experimentally measured ph-MTF values for the individual trenches in Fig.~\ref{fig:Exper2_limit} that match the simulated spatial frequencies and wrapping numbers. 
Circles (solid envelope) correspond to low-AR features ($\mathrm{AR}<0.5$); triangles (long-dashed envelope) correspond to intermediate-AR features ($0.5\le\mathrm{AR}<0.75$); and diamonds (short-dashed envelope) represent high-AR features ($\mathrm{AR}\ge 0.75$). 
Cyan, orange, and purple denote wrapping numbers 0, 1, and 2, respectively.}
   \label{fig:ExpPhMTF}
\end{figure}

Finally, we compare the simulated ph-MTF with experimental measurements from the patterned silicon sample in Fig.~\ref{fig:Exper2_limit}. 
For each resolvable spatial frequency, we identify trenches whose height range and wrapping number match the simulated conditions and compute an experimental ph-MTF value by normalizing the recovered phase modulation against a reference modulation measured on large-period trenches. 
Figure~\ref{fig:ExpPhMTF} plots these experimental points together with the simulated ph-MTF curves from Fig.~\ref{fig:LimitGeometry}(c). 
Marker color indicates the wrapping number (cyan, orange, and purple for 0, 1, and 2, respectively), while marker shape and the corresponding line style encode aspect ratio: circles with solid curves for $\mathrm{AR}<0.5$, triangles with long-dashed curves for $0.5\le\mathrm{AR}<0.75$, and diamonds with short-dashed curves for $\mathrm{AR}\ge 0.75$. 

For low-AR features ($\mathrm{AR}<0.5$), the experimental points cluster around the simulated envelopes, indicating that the ph-MTF remains a reliable indicator of phase-transfer fidelity in this regime. 
As AR increases, particularly for wrapped structures (orange and purple markers), the experimental ph-MTF shows a faster decay with spatial frequency and falls below the simulated curves. 
These deviations reflect additional sample- and system-specific factors not present in the idealized simulations, including finite feature size, imperfect boundary conditions, surface roughness, and small background phase or amplitude offsets arising from stray reflections and residual system aberrations.

Despite these effects, the overall behavior is consistent: phase transfer decreases monotonically with increasing spatial frequency, and larger AR structures exhibit earlier and more pronounced high-frequency roll-off. 
This qualitative agreement between simulation and experiment reinforces the conclusion that phase-transfer degradation is dominated by structural AR and associated wave-optical effects.

\section{Conclusion}
FPT enables large FOV, high lateral resolution, and quantitative topography reconstruction. Building on these capabilities, we introduced a dual-wavelength FPT framework that extends the unambiguous height range beyond the $\lambda/2$ limit of single-wavelength systems while preserving the lateral resolution defined by the synthetic NA. The method integrates a per-pixel wrapped-number search that enforces cross-wavelength height consistency without noise amplification, a digital-refocus procedure that co-registers multi-wavelength measurements without mechanical adjustment, and a global TV-regularized refinement with soft bound constraints that suppresses noise while preserving true discontinuities. Through simulations and experiments, we demonstrated accurate, spatially consistent surface reconstruction across an extended height range and quantified the role of structural AR in determining reconstruction accuracy.

At the same time, FPT fundamentally relies on a surface-like (quasi-2D) forward model in which all points remain illuminated and visible across the available illumination angles. Once a structure becomes too deep relative to its lateral opening, geometric occlusion and multi-path scattering violate this assumption and degrade phase-transfer fidelity. Thus, dual wavelengths extend the unambiguous height range but do not overcome the underlying geometric and wave-optical limits of the model. In the micron-scale regime examined here, high-frequency phase transfer breaks down well before the geometric visibility bound, indicating that partial shadowing, 3D scattering, and model inaccuracies—not the unwrapping step—set the practical aspect-ratio limit of FPT.

For context, coherence scanning interferometry and confocal reflectometry report much higher AR limits, but these results correspond to trench widths several tens to hundreds of micrometers~\cite{Ma2024_OE_NIR_CSI,Ma2024_ACSPhotonics_CSI_Comp,Jo2013_JOSK_WLI_TSV,OLT2024_QuasiConfocal}, far larger than the sub-micron structures targeted by FPT. At those scales, diffraction and 3D scattering are relatively weak and geometric shadowing becomes the primary constraint. In contrast, FPT is designed for micro- and sub-micron features where diffraction dominates, leading to substantially stricter AR limits.

Recent reflection-mode diffraction tomography employing multiple-scattering models~\cite{li2025rDTreflect,zhu2025rMSBP} have begun to directly address the 3D inverse problem for reflective, strongly scattering samples, thereby relaxing the surface-based assumptions inherent to FPT. These approaches enable full 3D reconstruction, but typically at the cost of substantially increased data requirements and computational complexity. Within this broader landscape, the proposed dual-wavelength FPT provides a complementary solution: it is computationally efficient and is well suited for wide-field, quasi-2D surfaces with large height variations but limited volumetric thickness. For the NA and wavelength configuration considered here, our AR-based analysis indicates that dual-wavelength FPT can be used reliably up to aspect ratios of about $\mathrm{AR}\approx 0.75$; beyond this regime, geometric visibility loss and 3D scattering progressively violate the surface-model assumptions and favor the use of full 3D reflection-mode diffraction-tomography-type approaches instead. We expect these AR-based criteria to serve as practical design guidelines for choosing between FPT and 3D models in future metrology systems, and a promising direction for future work is to integrate dual-/multi-wavelength strategies into reflection-mode diffraction tomography, combining an extended unambiguous height range with full 3D reconstruction for higher-AR and more complex microstructures.

\section*{Acknowledgments}
Funding was provided by Samsung Global Research Outreach and National Science Foundation (1846784). 
The authors thank the Boston University Shared Computing Cluster for providing the computational resources.  The authors thank Dr. Jiabei Zhu and Mitchell Gilmore for providing experimental and algorithmic assistance and advice. The authors also thank the rest of BU CISL group members for insightful discussions.

\nocite{*}

\bibliography{apssamp}

\end{document}